\documentclass{article}
\usepackage{spconf,amsmath,graphicx}

\usepackage{enumitem}
\setlist{nosep, leftmargin=14pt}

\usepackage{mwe} 
\usepackage{multirow}
\usepackage{amssymb}
\usepackage{amsmath}
\newcommand{\norm}[1]{\lVert#1\rVert}

\title{SAM-I2I: Unleash the Power of Segment Anything Model for Medical Image Translation}
\name{Jiayu Huo, S\'{e}bastien Ourselin, Rachel Sparks\thanks{Corresponding author: jiayu.huo@kcl.ac.uk}}
\address{School of Biomedical Engineering and Imaging Sciences (BMEIS), King's College London, London, UK}
\begin{document}
%
\maketitle
\begin{abstract}
Medical image translation is crucial for reducing the need for redundant and expensive multi-modal imaging in clinical field. However, current approaches based on Convolutional Neural Networks (CNNs) and Transformers often fail to capture fine-grain semantic features, resulting in suboptimal image quality. To address this challenge, we propose SAM-I2I, a novel image-to-image translation framework based on the Segment Anything Model 2 (SAM2). SAM-I2I utilizes a pre-trained image encoder to extract multiscale semantic features from the source image and a decoder, based on the mask unit attention module, to synthesize target modality images. Our experiments on multi-contrast MRI datasets demonstrate that SAM-I2I outperforms state-of-the-art methods, offering more efficient and accurate medical image translation.
\end{abstract}
\section{Introduction}
\label{sec:introduction}
Medical imaging is an essential tool for diagnosing various conditions, particularly in neurology and oncology~\cite{nie2018medical}. Multi-modality imaging, such as multi-modal MRI scans (e.g., T1, T2, and PD sequences), provides complementary information from different imaging sequences and can enhance diagnostic accuracy~\cite{yi2019generative}. Different MRI sequences visualize distinct tissue characteristics, helping in the detection of tumors, vascular malformations, or degenerative disease~\cite{motwani2013mr}. However, acquiring these different modalities incurs additional cost, including the need for extended patient time in the scanner. Furthermore, not all modalities may be available due to equipment limitations, patient conditions, or artifacts caused by motion during prolonged scanning sessions. These challenges drive a substantial demand for medical image translation techniques that enable the synthesis of one imaging modality from another, thereby reducing the necessity for multi-modal scans while preserving diagnostic fidelity.

Early medical image translation approaches were primarily based on the Generative Adversarial Networks (GANs), such as pix2pix~\cite{isola2017image} and CycleGAN~\cite{zhu2017unpaired}, which leverage adversarial training to generate realistic target modality images from an input source image. While GANs have promising results, they struggle with mode collapse. Besides, GAN-based methods can not generate realistic  pathological areas when training from scratch and the dataset scale is limited~\cite{dalmaz2022resvit}. Vision transformers have been applied to medical image translation since these models exploit self-attention mechanisms to capture long-range dependencies within features. For instance, ResViT~\cite{dalmaz2022resvit} is a hybrid architecture combining CNNs with transformer bottlenecks designed for cross-modality MRI synthesis. While each of these methods has contributed valuable insights challenges remain in achieving optimal performance across diverse clinical scenarios.

Vision foundation models, such as Contrastive Language-Image Pretraining (CLIP)~\cite{radford2021learning} and the Segment Anything Model (SAM)~\cite{kirillov2023segment}, have gained significant interest for their versatility and generalization ability across a wide range of tasks. CLIP aligns visual and textual representations by training on large-scale image-text pairs, enabling models to learn rich semantic relationships between the image and text embeddings. While CLIP is primarily used for tasks like image classification, its cross-modal learning ability has also inspired its use in medical contexts~\cite{zhao2023clip}. SAM is designed to handle segmentation tasks across any image domain by learning to detect and segment objects in a generalized manner. The image encoder in SAM captures multiscale semantic features, making it a promising tool for medical image translation, where the semantic features can support high-quality image synthesis.

In this paper, we designed a new framework SAM-I2I to perform cross-modality MRI translation leveraging the Segment Anything Model 2 (SAM2). Our approach utilizes the pre-trained Hiera image encoder used in SAM2 as the backbone model to provide multiscale semantic features. We design the decoder based on the mask unit attention module to effectively aggregate hierarchical features obtained from the backbone model to generate the target modality images. During the training stage, the weights of the backbone model are frozen and only the decoder is trained. Such a design preserves the representation capability of the backbone model trained on a large dataset. We conduct experiments on a publicly available multi-contrast MRI dataset and demonstrate the superiority of the SAM-I2I framework compared to other image-to-image translation methods.

\begin{figure*}[!htbp]
\centering
\includegraphics[width=\linewidth]{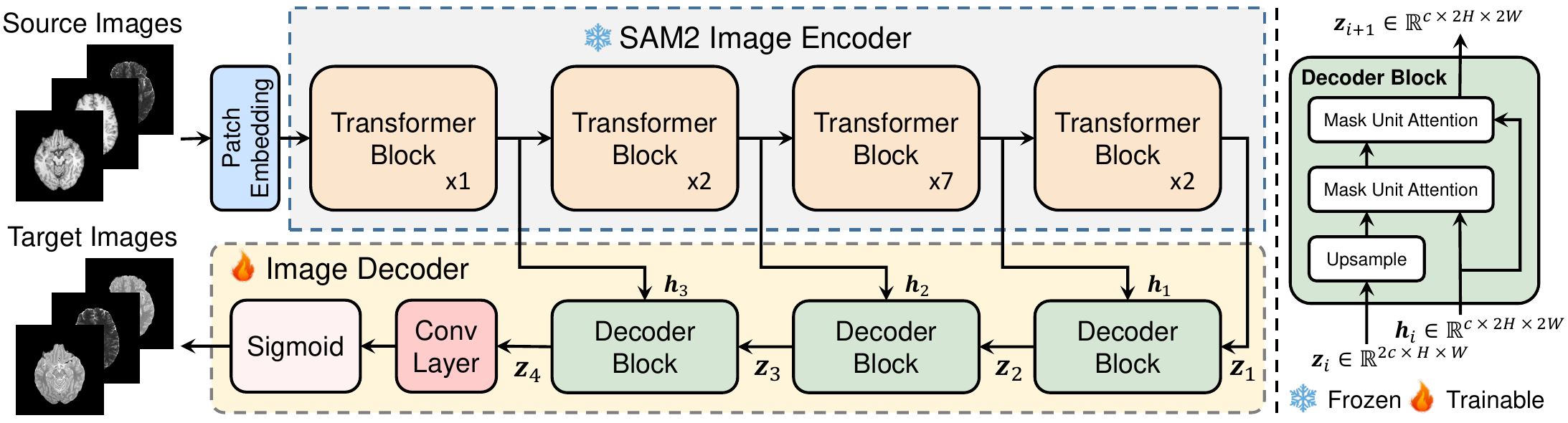}
\caption{Overview of our SAM-I2I framework. The pre-trained image encoder in SAM2 is the backbone model used to extract hierarchical features. The image decoder based on the mask unit attention module fuses multiscale features to generate the target modality images. We freeze the weights of the image encoder and only optimize the decoder during the training stage.}
\label{fig:main_fig}
\end{figure*}

\section{Methods}
\label{sec:methods}

The SAM-I2I architecture is shown in Figure~\ref{fig:main_fig}. SAM-I2I contains a pre-trained image encoder derived from SAM2 and an image decoder based on the mask unit attention module. The pre-trained image encoder is built upon Hiera~\cite{ryali2023hiera}, the hierarchical vision transformer used to produce multiscale features. For the image decoder, the mask unit attention module~\cite{ryali2023hiera} is used to decode features across resolutions to capture both global and fine-grained spatial details. Since the mask unit attention module separates features into small local windows, and computes the attention map within each window, the decoder can also reduces the computational cost.

\subsection{Hiera Image Encoder}
Hiera has four stages, each with a different number of transformer blocks to learn representative features across different scales. Unlike other vision transformers, which do not include pooling layers, Hiera progressively reduces the spatial size of features, allowing the model to retain more detailed spatial information. As the image encoder in SAM2, it has been trained on SA-1B~\cite{kirillov2023segment} and SA-V~\cite{ravi2024sam} datasets which contain over 1M images and 50K videos with 1.1B+ semantic masks. Training on such large datasets makes the Hiera image encoder have strong feature representations.

\subsection{Image Decoder with Mask Unit Attention}
We designed the image decoder to generate high-quality target modality images based on hierarchical features from the pre-trained image encoder. The proposed image decoder contains three blocks, each block takes two feature sets at different resolutions as the input, and fuses them with the mask unit attention module. Here we decided to not use the self-attention mechanism  as this can lead to a computational explosion when the feature spatial size is large (e.g., $256 \times 256$). The final convolution layer reduces the feature channel to one and is followed by a sigmoid activation to rescale values between 0 and 1.

In each mask unit attention module, features are first divided into several non-overlapping windows, then a multi-head attention is computed within each window. Such a design can effectively reduce the computational cost of the model, which makes it practical and efficient to apply the attention module to high-resolution features. In the model unit attention blocks, the window size for low-resolution features is larger than for high-resolution features, as global information is needed at lower resolutions to enable interactions across different regions. This helps the network to capture broader context within the image. Conversely, at higher resolutions the model focused on local information to generate fine-grained details for the final output.

\subsection{Objective Function}
During model training only the decoder weights are updated, the encoder weights are frozen to preserve the Heira model representation capability. To train the model, we use the L1 distance $\mathcal{L}_{img}$ to measure the difference between the output and target image intensities, as follows:
\begin{equation}
\mathcal{L}_{img} = \mathbb{E}_{x,y}[\norm{y-G(x)}_1].
\label{eq:loss_img}
\end{equation}
Here $x$ is the source domain image, $y$ is the target domain image, and $G$ is SAM-I2I. Only using $\mathcal{L}_{img}$ can result in over-smoothing~\cite{isola2017image}. Therefore, we introduce an adversarial loss $\mathcal{L}_{adv}$ defined as:
\begin{equation}
\mathcal{L}_{GAN} = \mathbb{E}_{y}[\log D(y)] + \mathbb{E}_{x}[\log (1-D(G(x))].
\label{eq:loss_gan}
\end{equation}
The discriminator $D$ uses the CLIP image encoder as a backbone model, to generate images with higher quality compared to training from scratch~\cite{kumari2022ensembling}. The final training objective function is the weighted combination:
\begin{equation}
\mathcal{L} = \lambda_{img} \mathcal{L}_{img} + \lambda_{GAN} \mathcal{L}_{GAN},
\label{eq:loss}
\end{equation}
where $\lambda_{img}$ and $\lambda_{GAN}$ are the weighted factors.

\begin{table*}[!htbp]
\caption{Quantitative performances of one-to-one image translation tasks on the IXI dataset using different image translation methods. We compare our approach to CNN and Transformer frameworks.}
\label{tab:main_result}
\centering
\fontsize{9}{11}\selectfont
\begin{tabular}{l|ccc|ccc|ccc|ccc}
\hline
\multirow{2}*{Model} &\multicolumn{3}{c|}{T1$\rightarrow$T2} &\multicolumn{3}{c}{T2$\rightarrow$T1} &\multicolumn{3}{c|}{T1$\rightarrow$PD} &\multicolumn{3}{c}{PD$\rightarrow$T1} \\
\cline{2-13}
&PSNR &SSIM &NRMSE &PSNR &SSIM &NRMSE &PSNR &SSIM &NRMSE &PSNR &SSIM &NRMSE \\
\hline
UNet     &27.00 &0.914 &0.2132 &29.09 &0.946 &0.0901 &27.06 &0.895 &0.1205 &29.03 &0.943 &0.0901 \\
Pix2Pix  &27.15 &0.920 &0.2100 &29.62 &0.946 &0.0841 &27.77 &0.914 &0.1110 &29.33 &0.945 &0.0870 \\
CycleGAN &27.60 &0.928 &0.1997 &29.77 &0.950 &0.0829 &27.97 &0.919 &0.1082 &29.80 &0.949 &0.0822 \\
ResViT   &27.89 &0.928 &0.1925 &30.19 &0.951 &0.0786 &28.20 &0.915 &0.1055 &30.20 &0.951 &0.0786 \\
Ours     &28.01 &0.931 &0.1904 &30.53 &0.955 &0.0757 &28.58 &0.926 &0.1013 &30.42 &0.953 &0.0767 \\
\hline
\end{tabular}
\end{table*}

\begin{table*}[!htbp]
\caption{Quantitative performances of one-to-one image translation tasks on the IXI dataset using different image encoders.}
\label{tab:different_backbone_result}
\centering
\fontsize{9}{11}\selectfont
\begin{tabular}{l|ccc|ccc|ccc|ccc}
\hline
\multirow{2}*{Encoder} &\multicolumn{3}{c|}{T1$\rightarrow$T2} &\multicolumn{3}{c}{T2$\rightarrow$T1} &\multicolumn{3}{c|}{T1$\rightarrow$PD} &\multicolumn{3}{c}{PD$\rightarrow$T1} \\
\cline{2-13}
&PSNR &SSIM &NRMSE &PSNR &SSIM &NRMSE &PSNR &SSIM &NRMSE &PSNR &SSIM &NRMSE \\
\hline
ResNet 101 &27.68 &0.928 &0.1974 &30.21 &0.953 &0.0781 &28.24 &0.923 &0.1052 &30.11 &0.952 &0.0795 \\
ViT-B-16   &27.31 &0.924 &0.2059 &30.09 &0.951 &0.0793 &28.09 &0.920 &0.1069 &30.02 &0.951 &0.0803 \\
Hiera      &28.01 &0.931 &0.1904 &30.53 &0.955 &0.0757 &28.58 &0.926 &0.1013 &30.42 &0.953 &0.0767 \\
\hline
\end{tabular}
\end{table*}

\section{Experimental Results}
\label{sec:results}
We evaluated the performance of SAM-I2I on the IXI dataset\footnote{https://brain-development.org/ixi-dataset/}. The IXI dataset contains 581 subjects each having a T1-, T2-, and PD-weighted MRI. The spatial resolution of each image is $0.94 \times 0.94 \times 1.2$ mm$^3$. We use 80\% of the subjects for the training set and the remaining subjects for the test set. For each MRI scan, we extract axial slices containing brain tissue to train the network. The following four one-to-one image translation experiments were conducted: T1 $\rightarrow$ T2, T2 $\rightarrow$ T1, T1 $\rightarrow$ PD, and PD $\rightarrow$ T1. For the model training, we use the Adam optimizer and set the learning rate to $2e^{-5}$. The batch size was set to 2 and all models were trained for 20 epochs. The $\lambda_{img}$ and $\lambda_{GAN}$ were set to 50 and 1.0, respectively. Synthetic images were evaluate with three metrics: Peak Signal-to-Noise Ratio (PSNR), Structural Similarity Index Measure (SSIM), and Normalized Root Mean Squared Error (NRMSE).

We compare SAM-I2I to the following CNN-based and transformer-based image translation models: UNet~\cite{xu2017200x}, Pix2Pix~\cite{isola2017image}, CycleGAN~\cite{zhu2017unpaired}, and ResViT~\cite{dalmaz2022resvit}. Quantitative results are shown in Table~\ref{tab:main_result}. Our approach SAM-I2I achieves the best performance for all tasks. Note ResViT constantly outperforms all CNN-based models, suggesting that applying a transformer block in the bottleneck has stronger feature learning capabilities for this dataset. The long-range modeling characteristics of the transformer block could be beneficial to cross-modality synthesis. However, SAM-I2I achieves an average 0.26 dB improvement in the PSNR metric on four cross-modality synthesis tasks compared to ResViT, which implies that effectively utilizing representative features provided by a powerful pre-trained image encoder is more suitable for cross-modality synthesis compared to training a model from scratch. Additionally, our results highlight that learning the transition function from other modalities to T1-weighted MRI appears to be more straightforward than the reverse task. We believe this is because T1-weighted images primarily contain general anatomical structure information, whereas converting from T1-weighted to other modalities requires generating more complex contrast and modality-specific details.

\begin{figure*}[!t]
\centering
\includegraphics[width=0.98\linewidth]{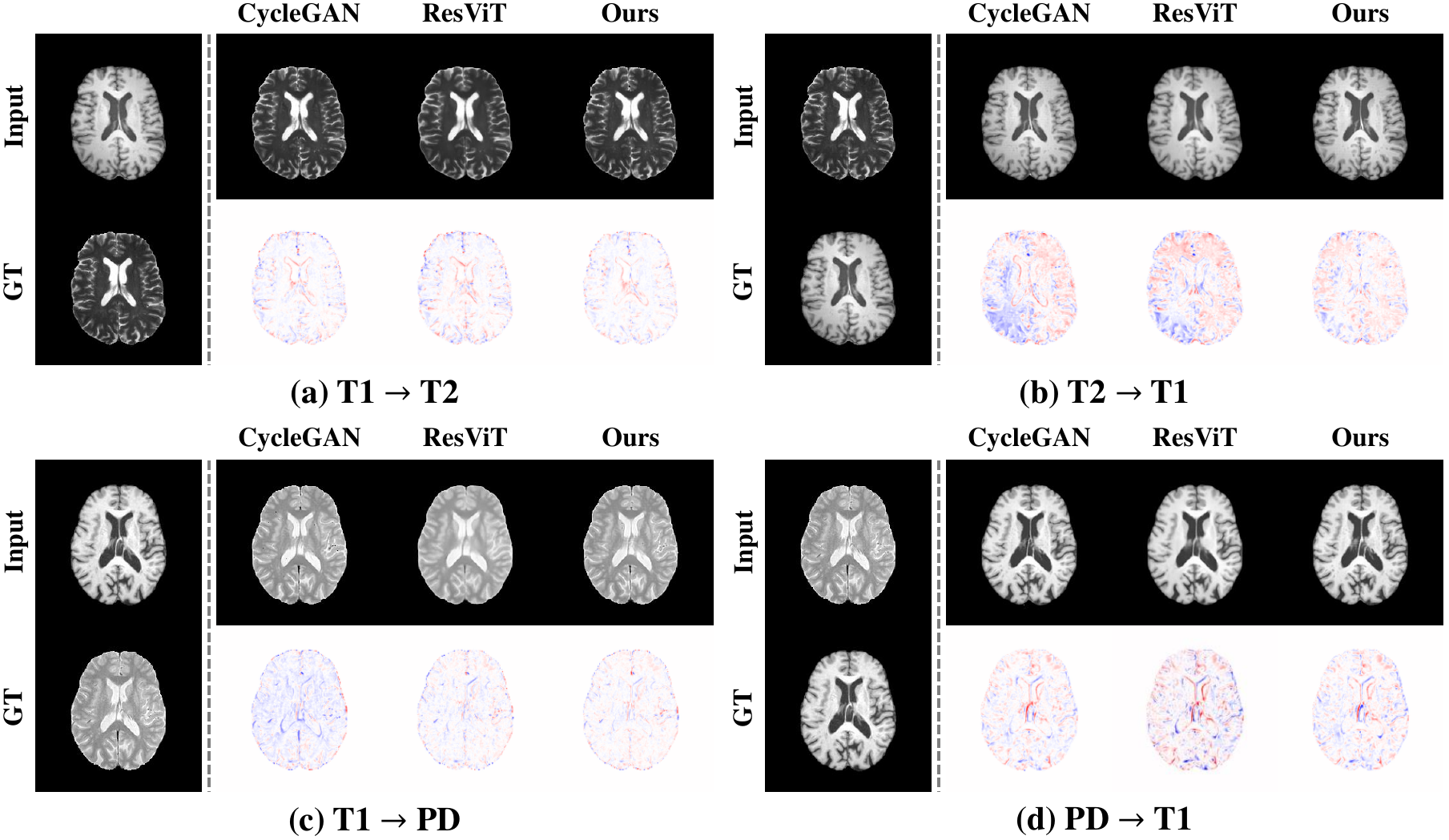}
\caption{Qualitative results of the image translation tasks. A $\rightarrow$ B indicates A is the source domain and B is the target domain. We provide source images, target images, and error maps for comparison.}
\label{fig:img2img_results}
\end{figure*}

Synthetic images generated by different methods are displayed in Figure~\ref{fig:img2img_results}. Outputs from our approach SAM-I2I show the most realistic textures and lowest error to the ground truth images. For the comparison methods, the synthetic target modality images tend to give overly smooth textures and unclear structure, especially for ResViT. Additionally, for the T1 $\rightarrow$ PD, the result generated by CycleGAN has some noise with the error map highlighting large differences. This indicates that CycleGAN is unable to model complex tissue contrasts effectively.

\begin{figure}[!htbp]
\centering
\includegraphics[width=\linewidth]{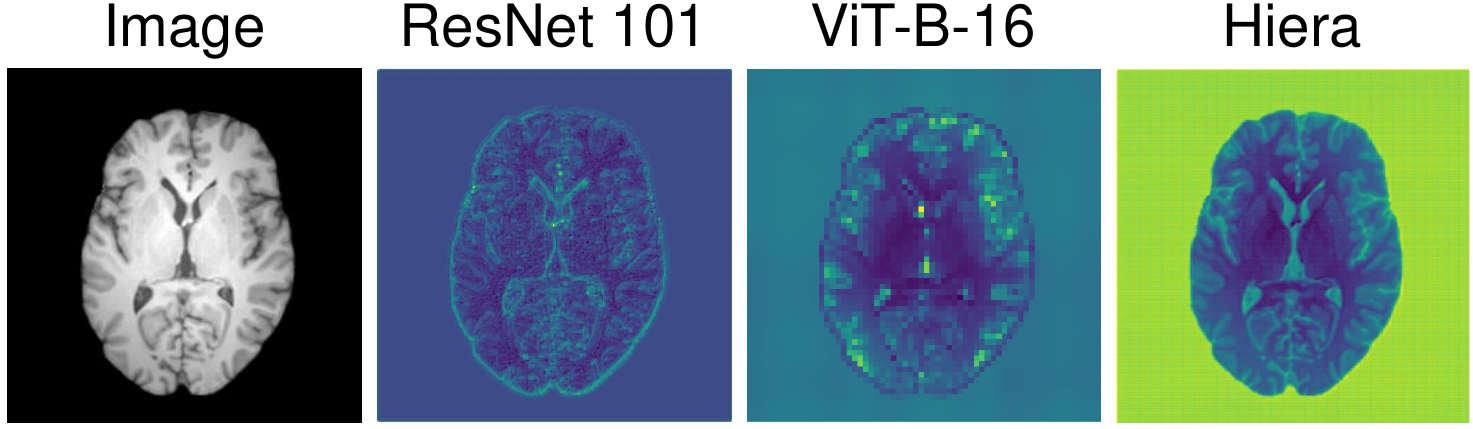}
\caption{The visualization of the first stage features from different pre-trained models. Features from the Hiera model contain more details compared to other models.}
\label{fig:all_model_features}
\end{figure}

We investigated how different backbone models affect the quality of the images synthesized. Quantitative results are shown in Table~\ref{tab:different_backbone_result}. The pre-trained weights of ResNet 101~\cite{he2016deep} and ViT-B-16~\cite{dosovitskiy2021an} are acquired from CLIP trained on the LAION-2B~\cite{schuhmann2022laion} dataset. ResNet 101 has four stages with different spatial resolutions, which is similar to the Hiera model. However, ResNet 101 does not contain any attention module. ViT-B-16 first partitions the image into several non-overlapping patches of $16 \times 16$ pixels and embeds each image patch into a feature vector. The feature vector set goes through a multi-head attention module for further refinement. 

ResNet 101 constantly yielded better results for all four tasks compared to ViT-B-16. This indicates multiscale features are pivotal for good quality image synthesis. ViT-B-16 was designed for image classification, which emphasizes semantic generalization ability while neglecting uninformative detailed local features. Therefore, ViT-B-16 may not be appropriate for image translation. The Hiera model maintains the ability to generate multiscale features and uses the self-attention mechanism to adaptively refine features at different levels. Hiera achieves better performance compared to ResNet 101 and ViT-B-16. The feature maps of the different backbone models are visually represented in Figure~\ref{fig:all_model_features}. The Hiera features preserve the most detail, while features extracted from ResNet 101 and ViT-B-16 lose the texture information or contain artifacts, respectively. This demonstrates the suitability of using the pre-trained Hiera model as the image encoder for image translation.

\section{Conclusion}
\label{sec:conclusion_and_discussion}
We propose the SAM-I2I framework for cross-modality MRI synthesis by leveraging the knowledge encoded in a vision foundation model (SAM2). The model utilizes the pre-trained Hiera image encoder from SAM2 as the backbone model to extract relevant hierarchical features. A trainable image decoder using the mask attention unit effectively aggregated previously extracted features and generated the target modality image. This design can reduce the computational cost and allows the network compute the attention map even when the spatial resolution of the feature maps is high. Quantitative results on the IXI dataset demonstrate the superiority of SAM-I2I compared to other image translation methods. Qualitative results indicate that SAM-I2I can generate target modality images with clear structures and realistic textures. Moreover, ablation studies using different image encoders demonstrate the suitability of the pre-trained Hiera model for medical image translation.

Currently, our evaluation of SAM-I2I has been limited to cross-modality MRI translation. Future work will be to apply SAM-I2I to a broader range of medical image translation tasks such as CT-to-MRI and PET-to-MRI. These applications would allow us to assess the effectiveness of SAM-I2I across other imaging modalities, addressing a wider scope of clinical needs. Additionally, we aim to investigate the feasibility of extending SAM-I2I to 3D volume image synthesis following the approach of the memory attention mechanism used in SAM2. 

\section{Acknowledgments}
This work was supported by Centre for Doctoral Training in Surgical and Interventional Engineering at King’s College London, and by core funding from the Wellcome/EPSRC Centre for Medical Engineering [WT203148/Z/16/Z]. The views expressed in this publication are those of the authors and not necessarily those of the Wellcome Trust. For the purpose of Open Access, the Author has applied a CC BY public copyright license to any Author Accepted Manuscript version arising from this submission.

\bibliographystyle{IEEEbib}
\bibliography{strings,refs}

\end{document}